\documentclass[12pt]{article}
\input{tcilatex}

\begin{document}

\begin{center}
\textbf{THE ANISOTROPIC XY MODEL ON THE 1D ALTERNATING SUPERLATTICE}
\end{center}

\textbf{\smallskip }

\begin{center}
F. F. Barbosa Filho$^{1}$, J. P. de Lima$^{1}$ and L. L. Gon\c{c}alves$^{2}$

\smallskip

$^{1}$ Departamento de F\'{i}sica da UFPi, Campus do Ininga, 64049-550
Teresina, Piau\'{i}, Brazil

$^{2}$Departamento de F\'{i}sica da UFC, Campus do Pici, C.P. 6030,
60451-970 Fortaleza, Cear\'{a}, Brazil

\smallskip

\textbf{Abstract}
\end{center}

The anisotropic XY-model in a transverse field $(s=1/2)$ on the
one-dimensional alternating superlattice (closed chain) is considered. The
solution of the model is obtained by introducing a generalized Jordan-Wigner
transformation which maps the system onto a non-interacting fermion gas. The
exact excitation spectrum is determined by reducing the problem to a
diagonalization of a block matrix, and it is shown that it is numerically
identical to the one obtained by using the approximate transfer matrix
method . The induced magnetization and the susceptibility $\chi ^{zz}$ are
determined as a function of the transverse field, and it is shown that, at $%
T=0,$ the susceptibility presents multiple singularities. It is also shown,
as expected, that this critical behaviour driven by transverse field belongs
to the same universality class of the model on the alternating chain.
\smallskip

\bigskip 

\smallskip 

\bigskip 

\bigskip 

\bigskip 


Corresponding author: J. P. de Lima, Departamento de F\'{i}sica da UFPi,

\ \ \ \ \ \ \ \ \ \ \ \ \ \ \ \ \ \ \ \ \ \ \ \ \ \ \ \ \ \ \ \ \ Campus \
do \nolinebreak Ininga,64049-550Teresina, Piau\'{i}, Brazil

\ \ \ \ \ \ \ \ \ \ \ \ \ \ \ \ \ \ \ \ \ \ \ \ \ \ \ \ \ \ \ \ \ (FAX:
+55-86-215-5560; e-mail: pimentel@mnnet.com.br)

\bigskip \pagebreak 

We have recently considered the isotropic XY model on an alternating
superlattice$^{1},$ which we have been able to solve exactly. Although the
model presents a critical behaviour, this behaviour is related to the
induced magnetization since, even at T=0, it does not present spontaneous
magnetization$^{2}.$ Therefore, in order to study the critical behaviour
induced by the transverse field when there is spontaneous order in the
ground state , we will consider the anisotropic XY model on the
one-dimensional alternating superlattice\- $(s=1/2)$. The Hamiltonian of the
system is given by

\begin{eqnarray}
\ H &=&-\frac{1}{2}\dsum\limits_{l=1}^{N}\left\{
\dsum\limits_{m=1}^{n_{A}}2h_{A}S_{l,m}^{A^{Z}}+\dsum%
\limits_{m=1}^{n_{B}}2h_{B}S_{l,m}^{B^{Z}}+\right.   \nonumber \\[0.1in]
&&  \nonumber \\
&&+\left\{ \dsum\limits_{m=1}^{n_{A}-1}J_{A}\left(
S_{l,m}^{A^{+}}S_{l,m+1}^{A^{-}}+\gamma
_{A}S_{l,m}^{A^{+}}S_{l,m+1}^{A^{+}}\right) \right. +  \nonumber \\
&&  \nonumber \\
&&+\dsum\limits_{m=1}^{n_{B}-1}J_{B}\left(
S_{l,m}^{B^{+}}S_{l,m+1}^{B^{-}}+\gamma
_{B}S_{l,m}^{B^{+}}S_{l,m+1}^{B^{+}}\right) +  \nonumber \\
&&  \nonumber \\
&&+J\left[ \left(
S_{l,n_{A}}^{A^{+}}S_{l,1}^{B^{-}}+S_{l,n_{B}}^{B^{+}}S_{l+1,1}^{A^{-}}%
\right) \right. +  \nonumber \\
&&  \nonumber \\
&&+\left. \left. \left. \gamma \left(
S_{l,n_{A}}^{A^{+}}S_{l,1}^{B^{+}}+S_{l,n_{B}}^{B^{+}}S_{l+1,1}^{A^{+}}%
\right) \right] \right\} +h.c.\ \right\} ,  \TCItag{(1)}
\end{eqnarray}
where $S^{\pm }=S^{X}\pm iS^{Y},$ $J$ and $\gamma $ are respectively the
exchange and anisotropy parameters between spins at the interfaces, $%
J_{A}(J_{B})$ and $\gamma _{A}(\gamma _{B})$ the exchange and anisotropy
parameters for spins within the subcell $A(B)$, and $h_{A}(h_{B})$ is the
transverse field in the medium $A(B)$. The spin operators can be expressed
in terms of fermion operators using the generalized Jordan-Wigner$^{3}$ 
\begin{equation}
S_{l,m}^{A^{+}}=\exp \left\{ i\pi
\dsum\limits_{n=1}^{l-1}\dsum\limits_{r=1}^{n_{A}}a_{n,r}^{\dag
}a_{n,r}+i\pi \dsum\limits_{r=1}^{m-1}a_{l,r}^{\dag }a_{l,r}+i\pi
\dsum\limits_{n=1}^{l-1}\dsum\limits_{r=1}^{n_{B}}b_{n,r}^{\dag
}b_{n,r}\right\} a_{l,m}^{\dag }\ ,  \tag{(2)}
\end{equation}
\begin{equation}
S_{l,m}^{B^{+}}=\exp \left\{ i\pi
\dsum\limits_{n=1}^{l}\dsum\limits_{r=1}^{n_{A}}a_{n,r}^{\dag }a_{n,r}+i\pi
\dsum\limits_{r=1}^{m-1}b_{l,r}^{\dag }b_{l,r}+i\pi
\dsum\limits_{n=1}^{l-1}\dsum\limits_{r=1}^{n_{B}}b_{n,r}^{\dag
}b_{n,r}\right\} b_{l,m}^{\dag }\ ,  \tag{(3)}
\end{equation}
and by introducing this transformation in Eq.(1), the Hamiltonian can be
written in the form

\begin{eqnarray}
\ H &=&-\dsum\limits_{l=1}^{N}\left\{
\dsum\limits_{m=1}^{n_{A}}h_{A}a_{l,m}^{\dag
}a_{l,m}+\dsum\limits_{m=1}^{n_{B}}h_{B}b_{l,m}^{\dag }b_{l,m}+\right.  
\nonumber \\[0.01in]
&&  \nonumber \\
\ + &&\dsum\limits_{m=1}^{n_{A}-1}\frac{J_{A}}{2}\left( a_{l,m}^{\dag
}a_{l,m+1}+\gamma _{A}a_{l,m}^{\dag }a_{l,m+1}^{\dag }+h.c.\right) + 
\nonumber \\[0.1in]
&&  \nonumber \\
+ &&\dsum\limits_{m=1}^{n_{B}-1}\frac{J_{B}}{2}\left( b_{l,m}^{\dag
}b_{l,m+1}+\gamma _{B}b_{l,m}^{\dag }b_{l,m+1}^{\dag }+h.c.\right) +\  
\nonumber \\
&&  \nonumber \\
\ + &&\frac{J}{2}\left. \left[ a_{l,n_{A}}^{\dag }b_{l,1}+b_{l,n_{B}}^{\dag
}a_{l+1,1}+\gamma \left( a_{l,n_{A}}^{\dag }b_{l,1}^{\dag
}+b_{l,n_{B}}^{\dag }a_{l+1,1}^{\dag }\right) +h.c.\right] \right\} + 
\nonumber \\
&&  \nonumber \\
+ &&\dfrac{N\left( n_{A}h_{A}+n_{B}h_{B}\right) }{2}\ \ ,  \TCItag{(4)}
\end{eqnarray}
where $a$'s e $b$'s are fermion operators, and we have neglected a boundary
term which, in the thermodynamic limit, will not give any contribution for
the static properties. Then, introducing in Eq. (4) the Fourier transforms

\begin{equation}
a(b)_{lj}=\frac{1}{\sqrt{N}}\sum_{Q}e^{-iQdl}A(B)_{Q,j},  \tag{(5)}
\end{equation}
where $Q=2\pi n/Nd,$ $n=1,2,3,...,N$ and $d=n_{A}+n_{B}$ is the size of the
cell, the Hamiltonian can be written as $H=-\sum_{Q}H_{Q},$ where

\begin{eqnarray}
H_{Q} &=&\sum_{m=1}^{n_{A}-1}h_{A}A_{Q,m}^{\dag
}A_{Q,m}+\sum_{m=1}^{n_{B}-1}h_{B}B_{Q,m}^{\dag }B_{Q,m}+  \nonumber \\
&&\sum_{m=1}^{n_{A}-1}\left( \frac{J_{A}}{2}A_{Q,m}^{\dag }A_{Q,m+1}+\gamma
_{A}A_{Q,m}^{\dag }A_{-Q,m+1}^{\dag }+h.c\right) +  \nonumber \\
&&\sum_{m=1}^{n_{B}-1}\left( \frac{J_{B}}{2}B_{Q,m}^{\dag }B_{Q,m+1}+\gamma
_{B}B_{Q,m}^{\dag }B_{-Q,m+1}^{\dag }+h.c\right) +  \nonumber \\
&&\frac{J}{2}\left[ A_{Q,n_{A}}^{\dag }B_{Q,1}+B_{Q,n_{B}}^{\dag
}A_{Q,1}e^{-iQd}\right. +  \nonumber \\
&&\left. \gamma \left( A_{Q,n_{A}}^{\dag }B_{-Q,1}^{\dag }+B_{Q,n_{B}}^{\dag
}A_{-Q,1}^{\dag }e^{-iQd}\right) +h.c.\right] +N\frac{\left(
n_{A}h_{A}+n_{B}h_{B}\right) }{2}.  \TCItag{(6)}
\end{eqnarray}

We can also write $H$ in the form

\begin{equation}
H=\sum_{Q}V_{Q}^{\dag }\mathbf{T}(Q)V_{Q},  \tag{(7)}
\end{equation}
where

\begin{equation}
V_{Q}^{\dag }=\left( A_{Q,1}^{\dag },...,A_{Q,n_{A}}^{\dag },B_{Q,1}^{\dag
},...,B_{Q,n_{B}}^{\dag
},A_{-Q,1},...,A_{-Q,n_{A}},B_{-Q,1},...,B_{-Q,n_{B}},\right)   \tag{(8)}
\end{equation}
and $\mathbf{T}(Q)$ is a matrix of dimension $2d\times 2d$ which represents
the quadratic Hamiltonian $H_{Q}.$ Since $[H_{Q1,}H_{Q2}]=0,$ the
diagonalization of the Hamiltonian is reduced to the diagonalization of $%
H_{Q}$, which can be written as a free fermion system 
\begin{equation}
H_{Q}=\sum_{k}E_{Qk}\xi _{Q,k}^{\dag }\xi _{Q,k},  \tag{(9)}
\end{equation}
where $E_{Qk}$ are the diagonal elements of $\ U_{Q}\mathbf{T}%
_{Q}U_{Q}^{\dag },$ and $U_{Q}$ is the unitary transformation which
diagonalizes $H_{Q},$which is determined numerically. Since we are in the
particle-hole representation, the energy spectrum is restricted to the
positive solutions, and this means that it presents $d$ branches. By using
the unitary transformation $U_{Q},$ we can express the fermion operators a's
and b' s in terms of the $\xi $' s. This allows us to obtain immediately the
induced magnetization per site and cell, which is defined as$^{1}$

\begin{equation}
\left\langle S_{cel}^{z}\right\rangle =\frac{1}{N\left( n_{A}+n_{B}\right) }%
\left[ \dsum\limits_{l=1}^{N}\left( \dsum\limits_{m=1}^{n_{A}}\left\langle
S_{l,m}^{A^{z}}\right\rangle +\dsum\limits_{m=1}^{n_{B}}\left\langle
S_{l,m}^{B^{z}}\right\rangle \right) \right] ,  \tag{(10)}
\end{equation}
where 
\begin{equation}
\left\langle S_{l,m}^{A^{z}}\right\rangle =\left\langle a_{l,m}^{\dag
}a_{l,m}\right\rangle -\frac{1}{2},\text{ \qquad }\left\langle
S_{l,m}^{B^{z}}\right\rangle =\left\langle b_{l,m}^{\dag
}b_{l,m}\right\rangle -\frac{1}{2},  \tag{(11)}
\end{equation}

which, at $T=0,$ can be written in the form

\begin{equation}
\left\langle S_{cel}^{z}\right\rangle =\frac{1}{N\left( n_{A}+n_{B}\right) }%
\left[ \dsum\limits_{m=1}^{n_{A}+n_{B}}\dsum\limits_{Q}\dsum%
\limits_{k(E_{Qk}>0)}u_{Q,km}^{\ast }u_{Q,km}\right] -\frac{1}{2}, 
\tag{(12)}
\end{equation}
where $u_{Q,km}$ are the matrix elements of the unitary transformation $%
U_{Q.}$

By making the identification $h_{A}\equiv h$ and $h_{B}\equiv \alpha h$, the
susceptibility $\chi ^{zz}$, can be obtained immediately from the previous
expression through the relation $\chi ^{zz}=\frac{d}{dh}\left\langle
S_{cel}^{z}\right\rangle ,$ which is determined numerically.

As an application of the results obtained, we have considered the case $%
n_{A}=n_{B}=2$ where the spectrum presents four branches $.$ For $J=1,$ $%
J_{A}=0.8,J_{B}=0.7,\gamma =0.1,$ $\gamma _{A}=\gamma _{B}=0.2$ and $\alpha
=2,$ there are four critical modes (vanish at $Qd=0$ $or$ $Qd=\pi )$ , which
define the four critical fields $h_{c1}\cong $ $0.01055,$ $h_{c2}\cong $ $%
0.36506$, $h_{c3\text{ }}\cong 0.54579$ and $h_{c4}\cong 0.65604$ . The
spectrum, as in the isotropic model$^{1},$ is numerically identical to the
one obtained by using the transfer matrix approximation$^{4}.$ It all should
be noted that, differently from the isotropic model$^{1}$, the number of
critical fields depends on the number of branches of the spectrum and on the
interaction parameters.

Since, in this case, there are four critical fields, we have four quantum
transitions induced by the transverse field, which are characterized by the
divergences of the susceptibility. This result is shown in Figure 1, where
the induced magnetization and the susceptibility, at $T=0,$ are presented as
a function of $h.$ Although for this case, the divergences of the
susceptibility are logarithmic $(log\mid h-h_{c}\mid )$, it is possible to
show, for special sets of parameters, that it can diverges as $\mid
h-h_{c}\mid ^{-1/2}.$ This implies that the model on the alternating
superlattice presents a parameter-dependent critical behaviour, and belongs
to the same universality class of the model on the alternating chain$^{5}.$

\textbf{Acknowledgements}. This work was partially financed by the Brazilian
agencies CNPq and Finep.

\bigskip

\textbf{References}

\smallskip

[1] J. P. de Lima and L. L. Gon\c{c}alves, J. Magn. Magn. Mater., \textbf{206%
}, 135 (1999)

[2] H. E. Lieb, T. Schultz and D. C. Mattis, Ann. Phys. \textbf{16}, 407
(1961)

[3] L. L. Gon\c{c}alves and J. P. de Lima, J. Magn. Magn. Mater., \textbf{%
140 -- 144}, 1606 (1995)

[4] L. L. Gon\c{c}alves and J. P. de Lima, J. Phys.: Condens. Matter,\textbf{%
\ 9}, 3447 (1997)

[5] J. H. H. Perk, H. W. Capel, and M. J. Zuilhof, Physica \textbf{81A}, 319
(1979)

\bigskip

\bigskip

\textbf{Figure captions.}

\smallskip

Figure 1. Induced magnetization (continous line) and the susceptibility $%
\chi ^{zz}$ (dotted line), for $n_{A}=n_{B}=2,$ $J=1,$ $J_{A}=0.8,$ $%
J_{B}=0.7,\gamma =0.1,$ $\gamma _{A}=\gamma _{B}=0.2,$ $\alpha =2,$at $T=0,$%
as a function of $h$ ($h_{c1}\cong $ $0.01055,$ $h_{c2}\cong $ $0.36506$, $%
h_{c3\text{ }}\cong 0.54579$ and $h_{c4}\cong 0.65604)$ .

\smallskip

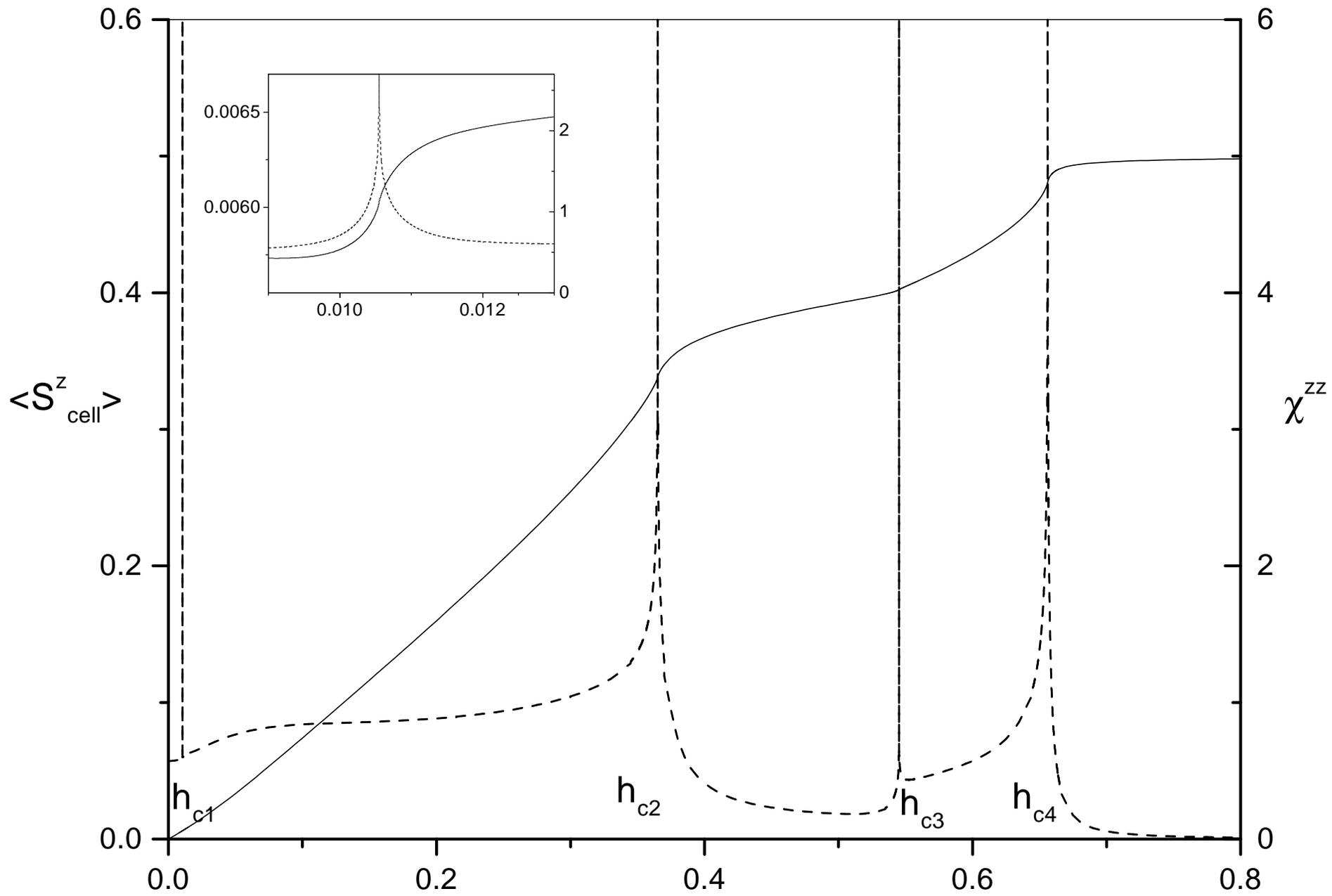
\end{document}